# Behavioural Analytics: Mathematics of the Mind

**R.O. Lane, H.M. State-Davey, C.J. Taylor, W.J. Holmes, R.A. Boon, M.D. Round**
QinetiQ, Malvern Technology Centre, St Andrews Road, Malvern, UK, WR14 3PS.

**Abstract**

Behavioural analytics provides insights into individual and crowd behaviour, enabling analysis of what previously happened and predictions for how people may be likely to act in the future. In defence and security, this analysis allows organisations to achieve tactical and strategic advantage through influence campaigns, a key counterpart to physical activities. Before action can be taken, online and real-world behaviour must be analysed to determine the level of threat. Huge data volumes mean that automated processes are required to attain an accurate understanding of risk. We describe the mathematical basis of technologies to analyse quotes in multiple languages. These include a Bayesian network to understand behavioural factors, state estimation algorithms for time series analysis, and machine learning algorithms for classification. We present results from studies of quotes in English, French, and Arabic, from anti-violence campaigners, politicians, extremists, and terrorists. The algorithms correctly identify extreme statements; and analysis at individual, group, and population levels detects both trends over time and sharp changes attributed to major geopolitical events. Group analysis shows that additional population characteristics can be determined, such as polarisation over particular issues and large-scale shifts in attitude. Finally, MP voting behaviour and statements from publically-available records are analysed to determine the level of correlation between what people say and what they do.

## 1. Introduction

Increasingly, malicious activities are carried out online rather than in the physical world. Current intelligence processes dictate that individuals who pose a potential threat to security are manually identified based on actions or statements they make, before being monitored in more detail. The pool of such people is increasingly diverse, and there are a range of ways they promote propaganda through the internet, using blogs, social media, and videos. Their online and real-world behaviour can be analysed to determine their threat level, so that appropriate action can be taken. However, large data volumes mean that automated processes are needed to assist analysts in understanding risk and to inform the actions of teams that endeavour to prevent harmful behaviour.

The first aspect of this paper proposes that analysts will use a language-independent system to track statements made by people over time and determine whether they are likely to become involved in extremism or terrorism. The use of such an automated system will allow analysts to examine a greater volume of data than is currently possible and to prioritise investigations to the most critical people. Once individuals or groups have been selected for further analysis there is a desire to understand their behaviour to enable target audience analysis (TAA), and potentially design behavioural-change interventions. Such interventions could be used to nudge people away from being involved in extremist or terrorist activities.

## 2. Behavioural Analytics Concepts and Architecture

### 2.1. Extremism and terrorism detection

The concept of tracking the descent of an individual into extremism and terrorism is illustrated in Figure 1. Each dot on the diagram represents a statement made by the person. The location of dots is a projection of a high-dimensional vector representation of each statement into two dimensions for visualisation purposes. Different regions of the diagram represent sentiments behind the statements. The hypothetical individual starts with 'normal' behaviour but over time becomes progressively more extreme and eventually goes on to carry out terrorist attacks.

The architecture of the proposed system is illustrated in Figure 2. The system comprises: online statement collation, a text-to-vector generator, a machine learning (ML) module, a tracker, and a visualisation system. The vector generator is fed with text data from online sources. In the training phase, the ML module uses manually-labelled data and the vectors to learn a model of terrorist, extremist, and 'normal' quotes. In the operational phase, a series of unlabelled statements can then be converted to vectors and tracked using the tracker. The ML module uses the output of the tracker to estimate whether or not the individual is currently or likely to be involved in extremism or terrorism. This information is communicated to the analyst using the visualisation system. More details of the system are given in Lane *et al.* (2021).

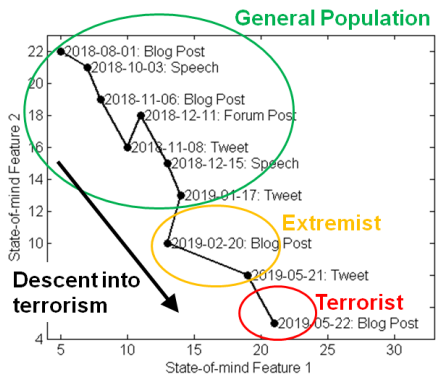 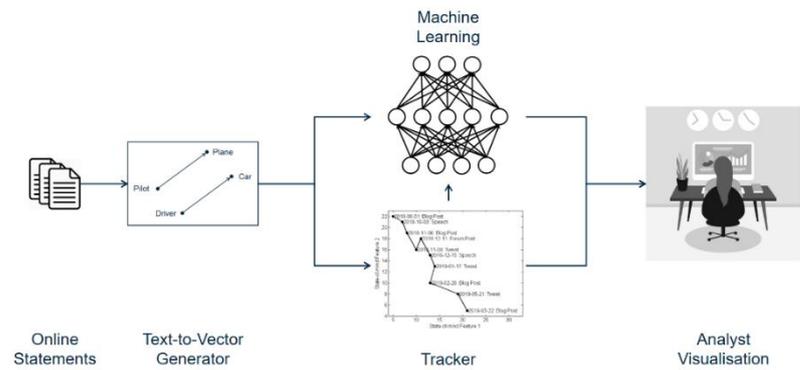

*Figure 1: Concept for tracking the descent into extremism and terrorism.*

*Figure 2: System architecture for tracking, classifying, and analysing the descent into extremism and terrorism.*

**2.2. Behavioural modelling**

Behavioural modelling is performed in this paper using a novel data-driven Bayesian network (BN) to understand which factors influence behaviour and to what extent. The initial structure of the model and relationship between components, including the key psychological and social factors that influence any given behaviour, is based on the capability-opportunity-motivation-behaviour (COM-B) model by Michie (2011), which provides expert knowledge for model structure definition. The COM-B model specifies the three factors (capability, opportunity, and motivation) that need to be present for a behavioural outcome. In the Bayesian network, these are hidden factors that in turn depend on the input factors of: skills (capability); social influence (opportunity); and emotion, attitude, subjective norm, and personality (motivation). The network architecture is shown in Figure 3. A further development of the work examines whether network structure can be estimated from the data alone.

**3. Data Sources**

**3.1. Introduction**

To demonstrate the behavioural analytics concept it is necessary to source pre-existing data that contain statements from a variety of individuals, where each person has made statements over a substantial time period, the quotes are time-stamped, and have favourable licensing conditions. Furthermore, to prove the language-agnostic approach, it is necessary to have data from a variety of languages. For this study, English, French, and Arabic data were used. Across all datasets, 6254 quotes were obtained and labelled. Quotes are analysed on two label-type axes: terrorism and Brexit. Although terrorism is the subject matter of interest, Brexit was found to be a useful topic to analyse. This is because voting records for UK MPs are available, allowing an analysis of the relationship between what people say and how they act. In addition to this, a much larger volume of unlabelled data has been analysed. The remainder of this section discusses data selection, including translation and labelling issues.

**3.2. English text and vote data**

Text data were collected from Wikiquote[1], a "free online compendium of sourced quotations from notable people and creative works in every language and translations of non-English quotes". Since the ultimate aim is to detect and analyse people from a general population that are likely to become involved in extremism or terrorism, the following groups of people were included for analysis:

- Terrorists (Islamist and Far Right);
- Extremists (Islamist and Far Right);
- Islamic anti-violence campaigners;
- UK and EU Politicians.

---

[1] https://www.wikiquote.org

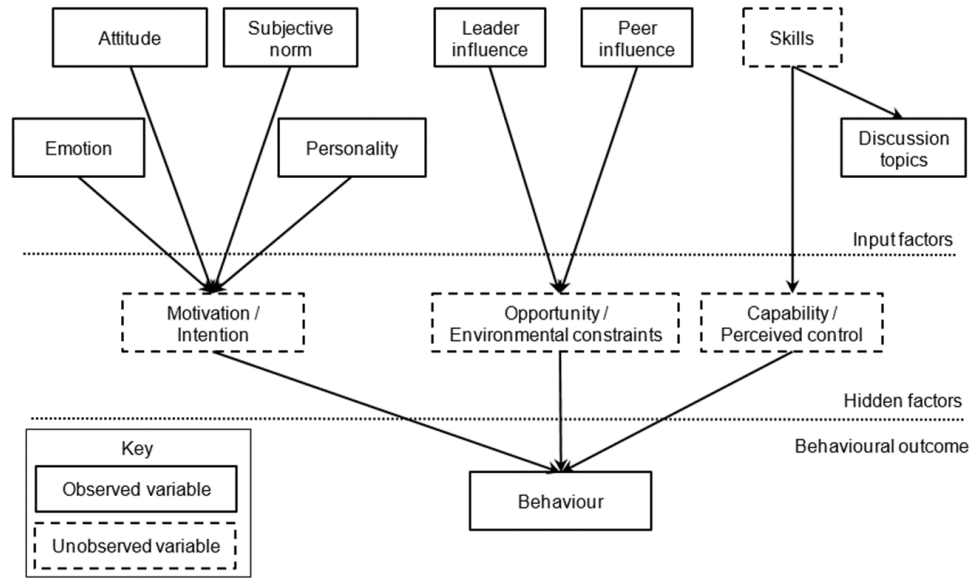

*Figure 3: High-level Bayesian network architecture for behavioural model, based on the COM-B model.*

Politicians occasionally talk about terrorism or extremism without being involved in it but mostly talk about other topics. We used them to act as a surrogate for the general population. Islamic anti-violence campaigners were included to ensure that bias is not inadvertently introduced to the analysis and that people who talk about Islamic topics are not incorrectly categorised. Each quote was labelled according to two label-type axes: terrorism and Brexit. On the terrorism axis, the possible labels were centrist (C), extremist (E), or terrorist (T). On the Brexit axis, the possible labels were: pro-hard Brexit (H), pro-soft Brexit (S), Brexit neutral (N), anti-Brexit (A), or other (O) (not to do with Brexit). More details about the Wikiquote data and the cleansing and labelling processes used during its generation are given in Lane *et al*. (2021). The final Wikiquote dataset included 1584 quotes. The number of labels for each Brexit category are: A: 176, N: 179, S: 174, H: 215, O: 840. The number of labels (not related to Brexit) for each terrorism category are: C: 682, E: 130, T: 28.

Although the Wikiquote dataset provides a good variety of people, the numbers of available quotes per person are limited. To use more data, a Hansard dataset was generated. Hansard is the official report of what is said in the UK Parliament. It provides access to a large and comprehensive body of data on parliamentary debates. Hansard data for the House of Commons were obtained from Zenodo[2] since much wrangling and cleaning of the data had already been done to facilitate analysis. In a first step to obtain Brexit-related quotes, all speeches made from 2015 through to 2019 were extracted. This covers the period containing general elections in 2015, 2017, and 2019. Quotes before the 2015 election and after the 2019 election are included, as the inclusion criteria were to collate data from the whole date range 2015-2019. This provided 310,396 parliamentary speeches. To prepare the data for analysis, cleaning was performed to correct and remove irrelevant and improperly formatted records, as well as remove quotes longer than 100 words. During this process, procedural quotes, such as announcements of reading a bill, were removed. Since the full dataset is too large to label comprehensively, a two-stage process was used. First, an active learning process, using the commercial off-the-shelf (COTS) Prodigy data labelling tool[3], was configured and used to select 1095 quotes likely to do with Brexit. Anonymised quotes were then comprehensively labelled by three independent human raters, to avoid bias, and the labels combined to give a single label for each quote. The number of labels for each category are: A: 646, N: 77, S: 204, H: 168, O: 0.

The labelled Hansard dataset has sufficient volume to conduct a number of useful experiments. However, real datasets are likely to be orders of magnitude larger and it would not be possible to label all the data manually. One practical task that analysts would be interested in performing with such data is the analysis of individual, group, or population attitudes, whether as a snapshot or as trends over time. This task can be aided by automated processes that either label the data or compute attitude from unlabelled data. To this end, a subset of the complete Hansard dataset was extracted to obtain quotes related to Brexit. The active learning model described above was used to assign a probability to each quote in the full dataset with 100 words or fewer as to whether it is Brexit-related. All quotes with a score higher than 85% were kept, along with those manually labelled as Brexit. The threshold was selected to provide a suitable balance between obtaining a high volume of quotes and selecting relevant ones. This resulted in 15,794 quotes being retained.

---

[2] https://zenodo.org/record/4066772, Evan Odell

[3] https://prodi.gy, Ines Montani and Matthew Honnibal

MP voting records relating to Brexit were collected from TheyWorkForYou[4], a parliamentary monitoring website. The website categorises voting by MPs on different issues. The information used for this study was on how MPs voted on "UK membership of the European Union (EU)". Voting records were collected for all MPs in the Hansard quote dataset. There were a total of 24 votes between 15th June 2016 and 29th March 2019. Each MP's votes were categorised as "for", "against", or "absent". On 8th June 2017 a general election took place where some MPs in the dataset lost or gained their seat. As a result, some MPs have fewer than 24 votes in total. For analysis, a "Pro EU membership" vote score for each MP was computed as the difference between for and against votes, measured as a proportion of the number of votes they were involved in. This gives each MP a vote score on a scale from -1 (against EU membership) to +1 (for EU membership). Records for 757 MPs were collected over the whole time period. MPs who had fewer than three quotes or an empty voting record were removed. This left a final dataset of 610 MPs.

### 3.3. French text

The French National Assembly is the lower chamber of the French parliament. The parliamentary cycle chosen for analysis started on 21st June 2017 and is known as the 15th legislature of the Fifth French Republic. The assembly publishes the text of its debates online[5] and is available for use under an open license modelled on the British Open Government License. Quotes were collected for the time period from 27th June 2017 to 12th December 2019. This date range matches as closely as possible the range of dates analysed in the Hansard data. The full dataset comprises 409,415 French language quotes from 577 deputies (members of the assembly) with associated metadata. Data cleansing included name normalisation, fixing formatting inconsistencies, and removing quotes longer than 100 words. No French behavioural data was obtained.

The Amazon Web Services (AWS) Translate service was used to translate all quotes into English. To avoid high translation costs, quotes were pre-filtered to those containing Brexit-related keywords in French, resulting in a pool of 4231 translated quotes. The same active learning process as used for the Hansard data to obtain Brexit-related quotes was applied to translations of the French assembly quotes. A single person rated 1267 quotes as either Brexit-related or not, with the model continuously being updated and presenting high-probability quotes to the user. Of these, 1132 were considered non-Brexit and automatically given the "other" label. The remaining 135 quotes were given to three people to independently assign one of the five labels: A, N, S, H, O. As with Hansard, labels from the three raters were combined to give a single label for each quote. The number of labels for each category are: A: 75, N: 45, S: 15, H: 0, O: 1132.

### 3.4. Arabic text

Due to difficulties in finding a single large source of timestamped Arabic quotes attributed to specific people and covering topics related to extremism or terrorism, multiple data sources were sought and used.

An initial manual assessment of the volume and quality of Arabic language quotes available on various news websites was carried out. Major news sites from the Middle East were considered (Al Jazeera, Al Riyadh, Al Arabiya, Al Masry Alyoum, and El Watan News), as well as western news sites containing Arabic pages: BBC News, Cable News Network (CNN), and France 24. A total of 63 quotes were obtained for a selection of well-known people including scholars, politicians, extremists, and terrorists. This process was useful to discover websites with good-quality information. However, it was a manually-intensive process that could only be performed by someone with knowledge of Arabic.

To obtain a larger volume of news media quotes, a semi-automated tool was developed to acquire quotes from BBC News Arabic and Al Jazeera. Quotes in the article text were automatically detected from the presence of quotation marks. AWS Translate was used to translate the quotes into English so that a user was presented with both English and Arabic quotes and the surrounding context. If accepted by the user as containing relevant information, each quote was recorded in the original Arabic. This tool enabled more rapid data gathering without requiring knowledge of Arabic. 149 quotes were obtained using this method, covering the same type of people as the fully manual analysis.

A survey of Arabic-speaking parliamentary websites was conducted. The Iraq Parliament Council of Representatives was found to publish data in a suitable format. Data were extracted from all 46 sittings (sessions) in the legislative year from 5th September 2020 to 31st March 2021. Data cleansing involved carrying out name normalisation, fixing formatting inconsistencies, and removing procedural quotes. The cleaned dataset contained 1995 quotes.

---

[4] https://www.theyworkforyou.com

[5] https://www.assemblee-nationale.fr/15/debats

The academic website Jihadology[6] is a "clearinghouse for jihadi primary source material", containing documents produced by extremist or terrorist organisations, including newsletters, organisations diagrams, and inventories. While the website provides useful information for manual analysis, much of it has been converted to a "safe" PDF format which prevents easy automated processing of the data. However, some information is available in HTML format allowing easier extraction. One section of the website contains reactions to the death of Osama bin Laden in 2011. Of these, 107 relevant and attributable quotes were manually extracted.

All quotes were translated into English using AWS Translate for human labelling. The manual-news, semi-automated-news, and Jihadology data sets were small enough that all 319 quotes were labelled according to terrorism label type by three independent raters. The Iraq Parliament dataset was too large for comprehensive labelling. Therefore an active learning approach was used to determine potentially interesting (extremist and terrorist) quotes to label. A single rater labelled Iraq Parliament quotes translated to English as extremist/terrorist or not. The model was continuously updated and high-probability quotes were presented to the user. Based on rating 1005 quotes across all Arabic data sets, no quotes were found in the Iraq Parliament data to be either extremist or terrorist. Based on this observation it was concluded that no such quotes exist in the dataset. Thus all Iraq Parliament quotes were given the label "centrist". The final Arabic dataset from all sources contained 2314 quotes. The number of labels for each category are: C: 2176, E: 117, T: 21.

## 4. Data Representation, Analysis, and Classification

### 4.1. Text to vector embeddings

To apply ML algorithms, the text must be converted into a form that captures the linguistic content in a way that is amenable to mathematical analysis. A variety of natural language processing (NLP) techniques have been developed that aim to provide a numerical representation that captures useful information about text content. A general approach to text representation that has become popular over the last decade is converting text into a semantic vector, also known as a thought vector, word embedding, or text embedding. This maps words or phrases into vectors of real numbers. This numerical representation is useful in multi-language domains, as transforming words into vectors is a language-independent technique capable of capturing semantic and syntactic similarity regardless of the language. Using language alignment techniques (Lample *et al.*, 2018), it is possible to label training text in one language and analyse new text in a different language, without requiring experts in each target language.

An extensive review and analysis of text-to-vector algorithms and software libraries has been conducted by the authors. An Euler diagram of the techniques analysed is shown in Figure 4. The outcome of the review was that the multilingual Universal Sentence Encoder (USE) by Google (Yang *et al.* 2019) is one of the most suitable tools. USE has the ability to process larger blocks of text than just words and has lower levels of bias compared to other techniques, when measured against sex, age, and names from different regions of the world (Cer *et al.* 2018). The encoder converts passages of text of any length to a 512-element vector. The review also determined that transformer-based tools (Vaswani *et al.*, 2017) should be compared to USE. A family of such tools is based on bidirectional encoder representations from transformers (BERT) (Devlin *et al.*, 2019). Separate monolingual models are available for each of the test languages, English, French, and Arabic. A multilingual version is also available for comparison. Results were generated for all these types of model.

### 4.2. Exploratory data analysis

Version 4 of the English USE[7] was used to compute embeddings of all labelled data from the Wikiquote and Hansard datasets, as well as the Arabic data translated into English. Linear Discriminant Analysis (LDA) was used to project the data down into two dimensions, using the labels to aid the projections. Results of the technique applied to the Arabic dataset translated into English, using the terrorism subject labels, are shown in Figure 5. The general (centrist) quotes form a dense cluster. Extremist quotes are more spread out and there is a small overlap with the centrist quotes. Terrorist quotes are mostly separated from the other categories. This pattern indicates that it should be possible to classify a reasonable proportion of quotes correctly. Similar results were found for the Wikiquote dataset (not shown).

---

[6] https://jihadology.net, a project by Aaron Y. Zelin.

[7] https://tfhub.dev/google/universal-sentence-encoder/4

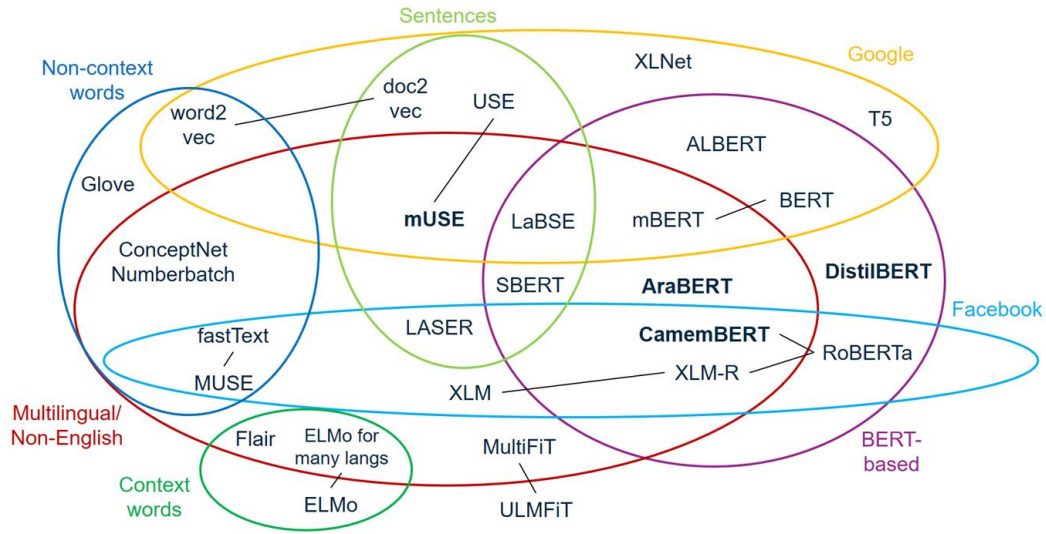

*Figure 4: Euler diagram of text embedding tools grouped by various characteristics.*

An equivalent LDA plot for Brexit-related quotes in the Hansard dataset is shown in Figure 6. For this plot, quotes in the soft-Brexit and hard-Brexit categories were merged to form a single pro-Brexit (P) category. There is reasonable separation of the anti, neutral, and pro clusters. The Hansard data (not shown) has slightly more overlap than the Wikiquote data. This may be because parliamentary debates included a lot of speeches criticising hard Brexit. In some cases these were labelled as pro-soft Brexit but would then have been grouped as part of the same pro-Brexit category. Such speeches have some similarity to purely anti-Brexit ones. Another notable feature of the data is that there is a higher degree of polarisation in Hansard, with fewer neutral quotes than Wikiquote. MPs in parliament often use speeches to criticise their political opponents, requiring more forceful argument than quotes made in the media that are aimed at a broader electorate.

In the above cases, the LDA model was trained using one set of quotes, and the same set of quotes were projected into 2D using that model. This can give an artificially simple view of the data, especially where the dimensionality of the input vectors is high compared to the number of quotes. When an LDA Brexit model is learnt from the Wikiquote data, but applied to the Hansard data, there is a very high degree of overlap in two dimensions (not shown). Although it is possible to obtain a good separation of clusters in 2D within each dataset, cross-application of the model results in clusters that are not well separated. This suggests there is a significant difference between the distributions of the two datasets. Manual analysis of the quote text does indeed show this to be the case. Parliamentary language is highly idiosyncratic in nature due both to rules of conduct and to the culture of speaking in a particular manner during debates. This demonstrates that for good classification of data it is necessary to train models on datasets that are valid for the target application.

Further analysis of the English, French, and Arabic datasets using LDA was carried out using embeddings computed with the multilingual version of USE. The patterns are largely similar to those found for English USE. Various models for the three languages, based on BERT, were used to project the data into 2D using LDA. These were DistilBERT for English (Sanh *et al.*, 2019), CamemBERT for French (Martin *et al.*, 2020), and XLM-Roberta for Arabic (Conneau *et al.*, 2020). XLM-Roberta is a multilingual model that was easier to use than the monolingual AraBERT model, which was investigated in initial work. The BERT projections generally had a higher level of overlap for different quote categories than when using either version of USE and were less useful for visualising separate clusters of the data.

### 4.3. Classification

The multilingual USE vectors were employed in conjunction with a Support Vector Machine (SVM) (Burgess, 1998) to develop a classifier that was applied to the data in a number of different experiments. As the quantity of data is moderate, especially for the extremist and terrorist categories, 10-fold cross validation (Arlot *et al.*, 2010) was used, ensuring that each quote contributed exactly once to the overall results while having no overlap between training and test data. Principle components analysis (PCA) dimensionality reduction was used to reduce noise in the data. The optimal number of PCA dimensions and parameters of the radial basis function (RBF) kernel used in the SVM were selected for each experiment based on performance in a validation set. The RBF kernel was chosen as it was found to outperform the linear kernel in all cases.

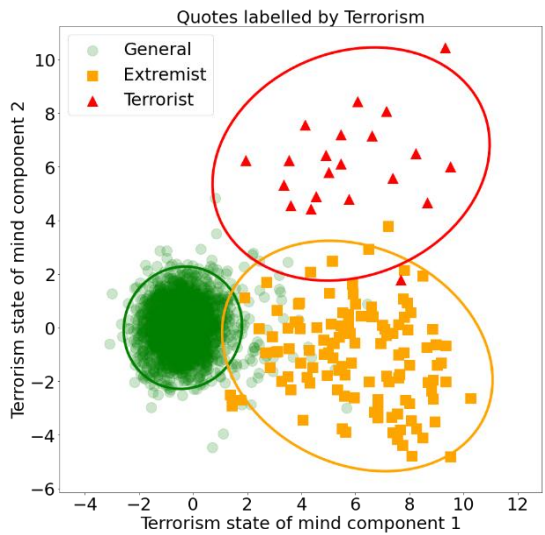

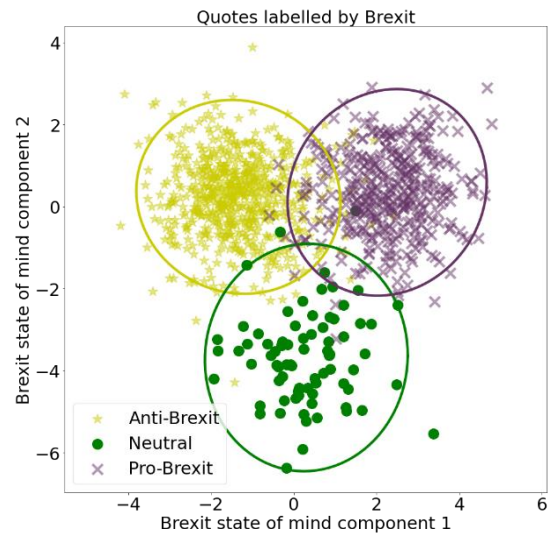

*Figure 5: Projection of terrorism quote vectors from English Universal Sentence Encoder 4 onto two axes using LDA, for Arabic data translated into English.*

*Figure 6: Projection of Brexit quote vectors from English Universal Sentence Encoder 4 onto two axes using LDA, for Hansard data (in English).*

The balanced-accuracy (BA) performance metric for these experiments is shown in Table 1. This metric is defined as the mean of the accuracy values computed for each class (also known as the macro-average), and is an appropriate metric for unbalanced datasets where there are many more or fewer examples of one class than the others.

| Dataset | Label set | Balanced accuracy |
| --- | --- | --- |
| English-Wikiquote | 2-way (Brexit, non-Brexit) | 94% |
| English-Wikiquote | 3-way (a, n, p) | 57% |
| English-Wikiquote | 4-way (a, n, s, h) | 47% |
| English-Hansard | 3-way (a, n, p) | 65% |
| English-Hansard | 4-way (a, n, s, h) | 54% |
| French | 2-way (Brexit, non-Brexit) | 85% |
| French | 3-way (a, n, s) | 48% |
| French translated to English | 3-way (a, n, s) | 54% |
| French | 4-way (a, n, s, o) | 52% |
| Arabic | 2-way (t, non-t) | 93% |
| Arabic | 2-way (c, e) | 95% |
| Arabic | 3-way (c, e, t) | 76% |
| Arabic translated to English | 3-way (c, e, t) | 78% |

*Table 1: Cross-validation classification results for multilingual USE with an SVM.*

The models for detecting whether or not a quote relates to Brexit performed well, with a BA of 85% for French parliament data and 94% for Wikiquote. The harder problem of three-way classification into anti, neutral, or pro-Brexit categories had a BA of 57% for Wikiquote and 65% for Hansard. Three-way classification of French parliament data into anti, neutral, or soft Brexit had a BA of 48%. However, interestingly, when all the French data was translated into English before training and classification, performance increased to 54%. This suggests that the English part of the model is better at characterising the world than the French part. This is likely due to the fact that vastly more English data than French was used to pre-train the USE model.

With Arabic data, the experiment to detect terrorist quotes achieved a performance of 93%. When classifying non-terrorist quotes into extremist or centrist, a performance of 95% was attained. However, the BA of the three-way classifier was lower, at 76%. Therefore it is better to split the problem into two binary problems than to attempt three-way classification in one go. It is interesting to note that breaking down multi-class problems into simpler ones has been shown to have advantages in other domains, e.g. image classification (Lane *et al*., 2023). When the Arabic data was translated into English before training and classification, three-way classifier performance increased to 78%. This result mirrors the translation results for the French data, in that translating to English improved results.

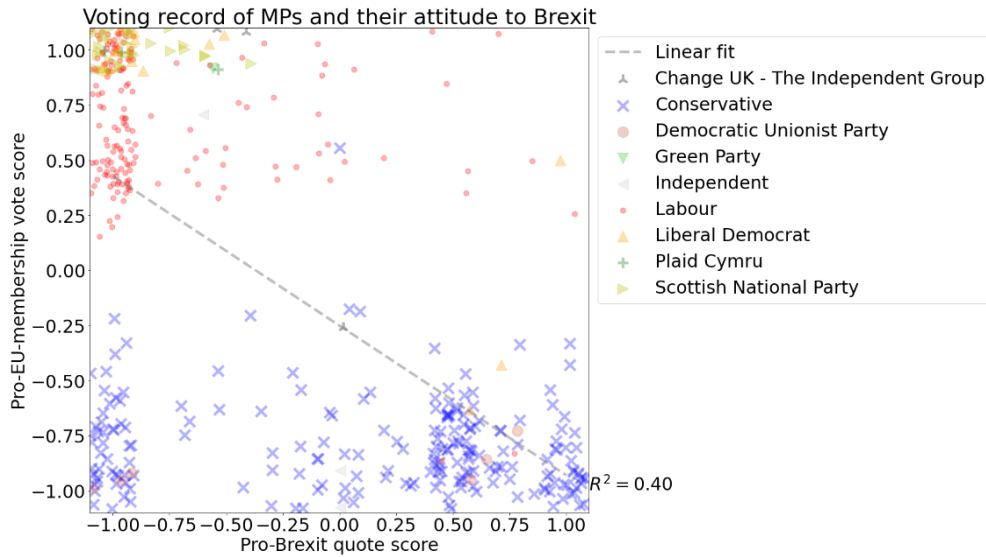

*Figure 7: Correlation between MP voting activity on EU membership and their attitude to Brexit. Data points have randomly been jittered by up to ±0.05 units on both axes to aid visualisation and prevent identical-location over-plotting.*

An additional three experiments compared the performance of the English USE model with multilingual USE, tested on the same English Wikiquote data. The multilingual USE performed better in all cases by 2% to 4% points (not shown). This suggests that additional abstract knowledge about the world is incorporated to the multilingual USE model, gained from the supplementary data in other languages used to train the model.

Similar experiments were carried out with the transformer-based BERT family of models, which don't require an SVM component. Due to the computational cost of training these models, only a subset of the experiments were run and they used a train-test split, rather than cross-validation. Multilingual USE performance was also assessed on the train-test split to provide a fair comparison. The USE BA was 4% to 10% lower for the train-test split than using cross-validation. This difference can be attributed to the fact that in cross-validation the training sets for each fold are larger, allowing a better model to be learnt. BERT generally performed worse than multilingual USE for the train-test experiments, with a BA of 2%, 13%, or 14% lower than USE for the three-class English-Hansard, French, and Arabic problems. However, CamemBERT, the French model, was 5% better at classifying quotes into Brexit or non-Brexit than multilingual USE.

The pattern of performances for the different experiments with BERT family models are similar to those for USE with SVM. Examination of the confusion matrices for BERT show that the classifiers were often significantly biased towards the class with most training data. It is thought that performance for the BERT models could be improved with a combination of hyper-parameter optimisation and techniques to deal with imbalanced datasets, such as oversampling and data augmentation (Wei *et al*., 2019).

**4.4. Correlation of attitude and behaviour**

Understanding the relationship between attitude and behaviour is important for supporting effective decision making where attitude data is used to predict potential behavioural action. The influence of attitude on behaviour was examined by comparing the Brexit quote data obtained from Hansard against MP voting records, as shown in Figure 7. The data exhibit a negative relationship between voting in favour of remaining in the EU and making pro-Brexit statements. This should generally be the expected direction of relationship, and it supports the hypothesis that statements relate to actual behaviour. However, there are a number of features of the relationship that merit further discussion.

Significant stratification of voting behaviour by party is evident, showing the influence of group membership on voting behaviour. While MP voting is usually "whipped" (i.e. MPs receive voting instructions from their party), Brexit is a topic that cuts across party lines, and divisions in both the main parties have been widely reported. Thus no assumptions were made regarding the specific effect of party membership on voting behaviour in the presence of relevant statements, prior to the analysis.

The pattern of attitudes and voting behaviour matches the known stance of the parties. For example, the largely pro-Brexit stance of the Conservative party is reflected in the lower pro-EU membership vote score for this party. The distribution of pro-Brexit quote scores also reflects the nature of the Conservative quotes which are mostly pro-hard Brexit but also include pro-soft Brexit and some anti-Brexit rhetoric. Members of the Labour party are shown to be primarily anti-Brexit, voting in favour of EU membership. Similarly, the Liberal Democrats and SNP members are shown to reflect a strong anti-Brexit stance in their parliamentary speeches and votes in favour of EU membership. These results show a significant correlation between attitude toward Brexit and actual voting behaviour, providing confidence that tracking statements made by individuals that are reflective of their attitudes and intentions can be used to predict likely behavioural outcomes. However, the strength of correlation between attitude and behaviour will differ depending on context and this should be considered in the analysis of other types of data.

## 5. Tracking and Prediction

### 5.1. Introduction

While it is of interest to classify either individual quotes or a person's general outlook based on their complete set of quotes, it is also particularly important to analyse trends over time. These trends can be used to determine whether someone's behaviour is relatively consistent, they are drifting towards extremism or terrorism, or a sudden change in behaviour has occurred. Sudden changes could be in reaction to an event or meeting a new influential person or group of people. Time-based analysis can be extended to groups or whole populations (Grindrod *et al*., 2011).

Statements made by individuals can be considered to be a noisy measurement of their state of mind. "Noisy" in this context is considered in the most general sense, including the fact that people can make successive statements on a wide variety of topics not necessarily related to each other. Using the framework developed in the previous section, this state of mind is represented as a high-dimensional vector in the same state-space as the vectors produced from the text of their statements. The state of mind can be tracked over time using standard tracking algorithms originally developed for sensor data, such as the Kalman filter (Welch *et al.*, 1995). The generated tracks produce not only an estimate of the current state of mind, but also its uncertainty. As more data are gathered, confidence in the current estimate and trajectory increases – the tracker estimate is more accurate than using a single statement. This information can be used either: to wait until a certain confidence is reached before alerting an analyst to a potential threat, or to better understand a person's characteristics during a forensic examination of the data. Future mind-states of the individuals under analysis can be estimated using the current trajectory, with near-term estimates being more accurate than longer-term ones. This estimation provides the capability to predict whether the tone of language being used, and consequently predicted behaviour, is becoming more extreme.

Lane *et al*. (2021) describe the mathematical details of a Kalman filter developed to track a 2D mind-state, based on the LDA projection of the high-dimensional vector representation of quotes. Here we describe an update to the algorithm to take into account how measurement noise depends on a person's state of mind rather than the measurement itself.

The tracker requires a measurement model for how statements z depend on a person's state of mind x. Based just on the labels for each quote, it has been observed that politicians in the data set rarely make extremist statements and never make terrorist ones. Extremists primarily make a mixture of centrist and extremist statements but tend not to make terrorist ones. Terrorists make a mixture of centrist, extremist, and terrorist statements. These observations about labels translate to distributions in the LDA projected data, which are visualised in Figure 8. The solid lines represent the distributions of quotes from the three centrist, extremist, and terrorist categories, when considered individually. The covariance matrix has been set the same for all three distributions, using a shared estimate. The shared estimate improves performance of the model, due to the limited number of extremist and terrorist quotes. Each person in the data set was assigned to one of the three categories, and all their quotes were grouped with quotes from other people in the same category. The dashed lines show the distribution of quotes for each of the centrist, extremist, and terrorist groups of people. As observed from the labels, centrists primarily make only centrist quotes (green dashed line), extremists make both extremist and centrist quotes (orange dashed line), and terrorists make all three kinds of quote (red dashed line).

Whilst Figure 8 shows the distribution of quotes given a type of person, it is possible to invert the model to estimate the type of person based on their quote, as shown in Figure 9. Based on an analysis of the diagram, if a person makes a terrorist quote then they are likely to be a terrorist (red dashed line). If a person makes an extremist quote then they are likely to be an extremist or terrorist (orange dashed line). If a person makes a centrist quote then they could belong to any of the three categories (green dashed line). However, so many of the centrist quotes are made by centrist people that the likelihood of someone who makes such a quote individually of being an extremist or terrorist is low.

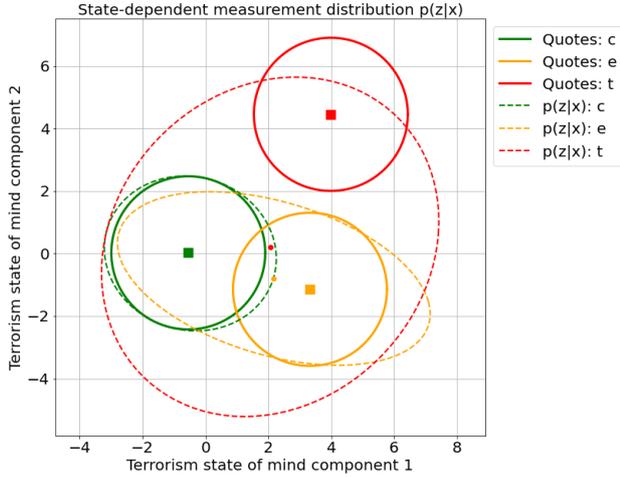
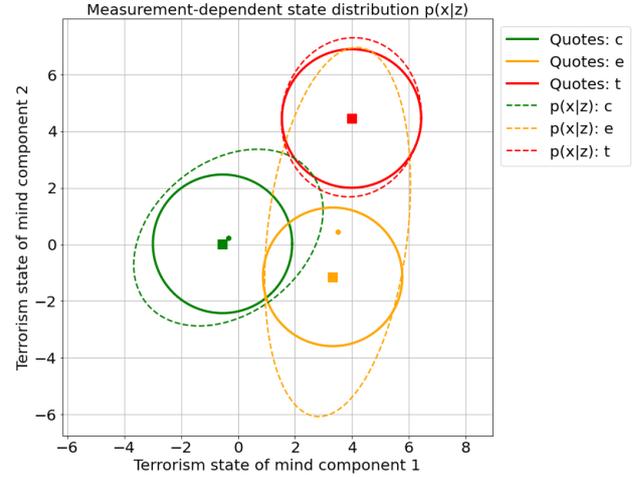

*Figure 8: Distribution of **measurement** vectors z for each category of quote (solid lines) and distribution of estimated **state-of-mind** vectors x for each category of **person** (dashed lines).*

*Figure 9: Distribution of **measurement** vectors z for each category of quote (solid lines) and distribution of estimated **state-of-mind** vectors x for each category of **quote** (dashed lines).*

The measurement model was used in conjunction with a transition model, which characterises how a person's state of mind changes over time, to build a tracker. The output of the tracker is an estimate of a person's time-varying state of mind along with the uncertainty of the estimate. This qualitative discussion is quantified using statistics in the following section.

### 5.2. Tracker mathematical details

A person's state of mind $\mathbf{x}$ is postulated to change over time $t$ according to a nearly constant velocity model $\mathbf{x}_{t+\Delta T} = \mathbf{F}\mathbf{x}_t + \mathbf{w}$, where $\mathbf{x} = [x_1, x_1', x_2, x_2']^T$, $\mathbf{F}$ is the transition matrix, and $\mathbf{Q}$ is the covariance of the zero-mean normally distributed process noise $\mathbf{w} \sim N(\mathbf{0}, \mathbf{Q})$. Each two-dimensional quote vector $\mathbf{z}_t$ is modelled as being a noisy measurement of the underlying state of mind at that time via $\mathbf{z}_t = \mathbf{H}\mathbf{x}_t + \mathbf{v}_t$. The matrix $\mathbf{H}$ selects the $x_1$ and $x_2$ variables from $\mathbf{x}$.

The model described so far is that same as in Lane *et al.* (2021), which also specifies the structure of the $\mathbf{F}$, $\mathbf{Q}$, and $\mathbf{H}$ matrices. The remainder of the model definition described below differs from that by using state-dependent measurement noise rather than a measurement-dependent measurement noise. Statistics for different categories of statement and state-of-mind states have also been updated to take into account the larger datasets studied here.

The noise vector $\mathbf{v}_t$ has a zero-mean normal distribution with covariance $\mathbf{R_x}$ that depends on the current state. The reason for using state-dependent noise is discussed as follows. Based on proportions in the dataset, the probability of making a statement of type $s \in \{c, e, t\}$ given the person is in a state-of-mind type $k \in \{c, e, t\}$ is encoded in the matrix (1), where the columns sum to unity. The probability of the person being in mind-state $k$ given they made a statement $s$ is shown in (2), where the rows sum to unity.

$$p(s|k) = \begin{bmatrix} s \backslash k & c & e & t \\ c & 0.983 & 0.293 & 0.356 \\ e & 0.168 & 0.707 & 0.475 \\ t & 0 & 0 & 0.169 \end{bmatrix} \quad (1) \quad p(k|s) = \begin{bmatrix} s \backslash k & c & e & t \\ c & 0.926 & 0.014 & 0.060 \\ e & 0.122 & 0.259 & 0.620 \\ t & 0 & 0 & 1 \end{bmatrix} \quad (2)$$

The marginal probabilities of making a statement $s$ or being in mind-state $k$ are shown in (3) and (4).

$$p(s) = \begin{bmatrix} c & e & t \\ 0.863 & 0.112 & 0.025 \end{bmatrix} \quad (3) \quad p(k) = \begin{bmatrix} c & e & t \\ 0.813 & 0.04 & 0.146 \end{bmatrix} \quad (4)$$

The Kalman filter requires knowledge of $p(\mathbf{z}|\mathbf{x})$. This is derived in equations (5) to (10). The only assumptions are in (7), where it is assumed the measurement $\mathbf{z}$ is independent of $k$ and $\mathbf{x}$ given the statement type $s$, and (9), where it is assumed the statement type $s$ is independent of the mind-state type $k$ given $\mathbf{x}$.

$$p(\mathbf{z}|\mathbf{x}) = \sum_{s,k} p(\mathbf{z}, s, k|\mathbf{x}) \quad (5)$$

7th IMA Conference on Mathematics in Defence and Security, London, UK, 7th September 2023

11$$p(\mathbf{z}|\mathbf{x}) = \sum_{s,k} p(\mathbf{z}|s,k,\mathbf{x})\, p(s,k|\mathbf{x}) \tag{6}$$

$$p(\mathbf{z}|\mathbf{x}) = \sum_{s,k} p(\mathbf{z}|s)\, p(s,k|\mathbf{x}) \tag{7}$$

$$p(\mathbf{z}|\mathbf{x}) = \sum_{s,k} p(\mathbf{z}|s)\, p(s|k,x) p(k|\mathbf{x}) \tag{8}$$

$$p(\mathbf{z}|\mathbf{x}) = \sum_{s,k} p(\mathbf{z}|s)\, p(s|x) p(k|\mathbf{x}) \tag{9}$$

$$p(\mathbf{z}|\mathbf{x}) \propto \sum_{s,k} p(\mathbf{z}|s)\, p(\mathbf{x}|s) p(s) p(\mathbf{x}|k) p(k) \tag{10}$$

The quantities on the right hand side of (10) are all known. Functions $p(s)$ and $p(k)$ are given in (3) and (4). The distribution $p(\mathbf{z}|s)$ is the distribution of quote vectors for a given quote type and is represented by the solid lines in Figure 8 and Figure 9. The distribution $p(\mathbf{x}|k)$ is the distribution of quote vectors for a given mind-state type and is represented by the dashed lines in Figure 8. The distribution $p(\mathbf{x}|s)$ is the distribution of mind-state vectors for a given quote type and is represented by the dashed lines in Figure 9. Mathematically, these normal distributions are given in (11) to (13), where the parameters are estimated from the full dataset of all people.

$$p(\mathbf{z}|s) = N(\mathbf{z}|\boldsymbol{\mu}_\mathbf{z}^s, \boldsymbol{\Sigma}_\mathbf{z}^s) \tag{11}$$
$$p(\mathbf{x}|k) = N(\mathbf{x}|\boldsymbol{\mu}_\mathbf{x}^k, \boldsymbol{\Sigma}_\mathbf{x}^k) \tag{12}$$
$$p(\mathbf{x}|s) = N(\mathbf{x}|\boldsymbol{\mu}_\mathbf{x}^s, \boldsymbol{\Sigma}_\mathbf{x}^s) \tag{13}$$

Equation (10) is a Gaussian mixture distribution over $\mathbf{z}$. For use in the Kalman filter, this is approximated in (14) by a single Gaussian $N(\mathbf{z}|\mathbf{x}, \mathbf{R_x})$, where $\mathbf{R_x}$ is the covariance of (10) computed via mixture reduction (Salmond, 2009).

$$p(\mathbf{z}|\mathbf{x}) \approx N(\mathbf{z}|\mathbf{Hx}, \mathbf{R_x}) \tag{14}$$

The above Kalman filter was applied to quotes made by various people. The process noise variance $\sigma^2$ was set to $0.1^2$. Independent prior distributions $p(x_1) = p(x_2) = N(0, 4^2)$ and $p(x_1') = p(x_2') = N(0, 0.3^2)$ were used. The priors for $x_1, x_2$ were based on the distribution of all quotes for all people. Noise variance $\sigma^2$ and priors for $x_1', x_2'$ were set based on manual data analysis and judgement.

**5.3. Tracker results**

The tracker has been applied to the combination of English quotes and Arabic quotes translated into English made by Osama bin Laden and Abu Bakr al-Baghdadi. The tracker output in terms of a two-dimensional state-of-mind for bin Laden is shown in Figure 10 as the solid line, with colour representing the date of the quote. For visualisation purposes, the track is overlaid onto the decision regions of a simple centrist/extremist/terrorist linear classifier trained on the full dataset. The time series of quote vectors is also shown, using a semi-transparent line with the same colour scheme as the track. Initial quotes made by bin Laden are considered extremist but not terrorist. He then went on to make more terrorist statements for a period of time after 2001 before returning to a less intense terrorist standpoint. His final quote in the dataset in 2011 moved the track further back into terrorist territory.

The effect of changing the measurement model from being measurement-dependent to state-dependent can be seen by comparing Figure 12 of Lane *et al.* (2021) to Figure 10 here. The previous measurement-dependent model allows the estimate of bin Laden's state of mind to reduce from terrorist to extremist at the end of the trajectory due to just two non-terrorist statements. This does not happen with the new state-dependent model in Figure 10, which would require a much larger proportion of someone's later statements to be non-terrorist for their assessment to be downgraded. This better reflects the real world, in which terrorists often make non-terrorist statements.

The same type of plot is shown for Abu Bakr al-Baghdadi in Figure 11. Initial quotes were considered extreme but not terrorist. However, sometime after al-Baghdadi's group broke with al-Qaeda in 2014 and renamed itself to Islamic State, al-Baghdadi started making more terrorist statements and the track increases in state-of-mind component 2 (moving up vertically in the 2D plot).

7th IMA Conference on Mathematics in Defence and Security, London, UK, 7th September 2023

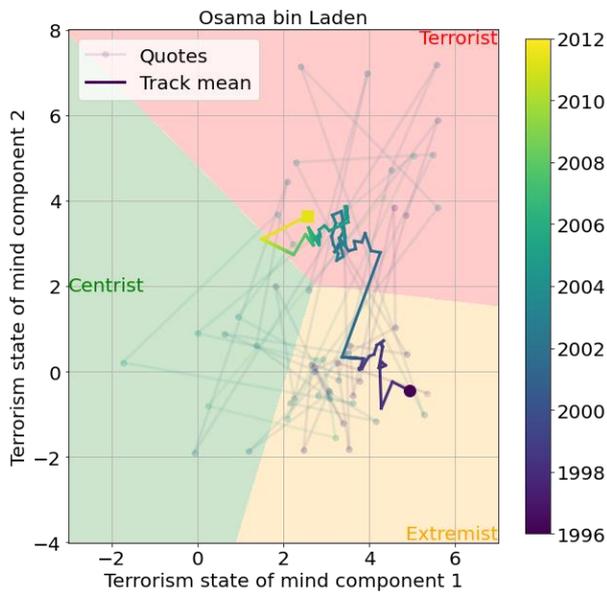
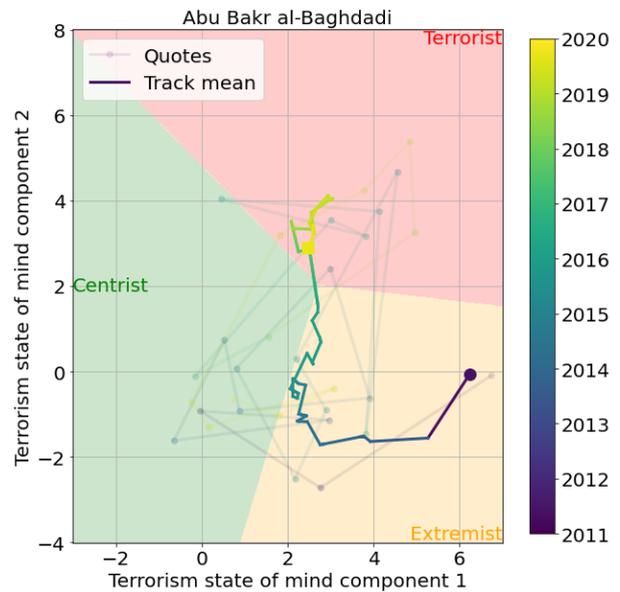

*Figure 10: Osama bin Laden quotes and state-of-mind track in 2D, overlaid onto linear classifier regions.*

*Figure 11: Abu Bakr al-Baghdadi quotes and state-of-mind track in 2D, overlaid onto linear classifier regions.*

## 6. Bayesian Network Behaviour Model

### 6.1. Introduction

This section describes the use of Bayesian networks to support the development of a practical data-driven model of human behaviour for more intelligent target audience analysis.

BNs are a popular and effective method for modelling uncertain and complex domains, such as behavioural change. They also provide a robust and mathematically coherent framework for the analysis of uncertain and complex problems. BNs provide a number of advantages that make them useful for application to the defence and security environment: they can handle missing or sparse data, they allow data to be combined with domain knowledge (Uusitalo, 2007), and they can show good prediction accuracy even with small sample sizes (Kontkanen *et al*. 1997). Developing BN models also forces researchers to think about how to articulate the subject under study in the form of a model, in contrast to an opaque-box approach. This encourages models to be based on sound scientific foundations, and ensures the model captures all factors that are necessary and sufficient for understanding the problem. Importantly, for understanding behaviour influence, BNs allow quantification of:

- The statistical relationship between influencing behavioural factors;
- The uncertainty in numerical estimates of factors, based on limited data;
- The probability of behaviours occurring; and
- How the likelihood of the desired behavioural outcome depends on targeted changes to the inputs.

BNs provide a measure of uncertainty using probability. "Beliefs" about the values of variables are expressed as probability distributions (i.e. a mathematical function that provides the probabilities of occurrence of different possible outcomes of an event). The higher the uncertainty, the wider the probability distribution will be. As information increases, knowledge of the true value of the variable increases, thus reducing uncertainty and narrowing the probability distribution represented in the BN. The probabilistic representation of uncertainty offers the advantage of preventing overconfidence in the strength of responses obtained when manipulating parts of the system or network, such as conducting "What If?" (sensitivity) analysis. An example of BNs being used for behavioural modelling is given by Lane *et al*. (2010), who use a BN to characterise the behaviour of ship captains based on measured ship trajectories.

The remainder of this section describes: variables used in the present analysis; the development of expert-specified, data-driven, and hybrid model structures; and a performance analysis of each network.

## 6.2. Variable definition

The English Hansard quote and vote dataset described in section 3.2 was used to test the proof-of-concept of using a BN to model human behaviour. Figure 3 shows the high-level concept of the Bayesian network. For this network to be instantiated, each of the input factors must be operationalised in a measurement model, and were implemented as follows.

Attitude, or more specifically attitude toward a behaviour, refers to the general feeling of favourableness or unfavourableness for that behaviour. Attitude was measured as the percentage of times an individual made a pro-Brexit quote out of all their Brexit-related quotes, based on the labels assigned by the reviewers.

Emotion is an affective attitude and refers to a person's emotional reaction (positive or negative) to performing the behaviour. Emotion was measured using the tool Synsketch (Krcadinac *et al.*, 2013), which produces a score for each sentence for the big six emotion categories: happiness, sadness, anger, fear, disgust, and surprise. The score for an individual in each of the categories was the average over all the sentences in all their quotes.

Subjective norm is the perceived opinion of other people in relation to the behaviour in question. In our voting use case we took the social group for the subjective norm to be the party of which each MP was a member. The norm refers to whether the expected behaviour of the social group is to vote for or against Brexit. It was measured as the mean probability of political party peers being in favour or Brexit. Thus each MP in the same party was assigned the same subjective norm score.

Personality was measured using the five-factor model (FFM) or "Big 5". FFM describes human personality using five dimensions: openness to experience, conscientiousness, extraversion, agreeableness, and neuroticism (OCEAN), upon which individuals are expected to vary on a continuum from low to high. Previous research showed FFM incorporates most known personality traits and, therefore, represents the basic structure underlying human personality (Goldberg, 1993). The FFM was measured using an automated machine learning assessment developed by the authors based on Empath features extracted from language text. Empath is a text analysis tool with a lexicon that covers a broad range of topics, and many categories such as violence and social media (Fast *et. al.*, 2016).

Leader influence was measured as the probability of the party leader being in favour of Brexit, and peer influence as the probability of each person being in favour of Brexit. The probabilities were estimated from the percentage of times each individual made a pro-Brexit quote out of all their Brexit-related quotes – the "attitude" variable described above. For peer influence, a trust network was defined where trust relationships between each pair of people are binary values (trusted or not trusted) based on party membership. In this simplified model, all MPs trust themselves and their party colleagues, and all other MPs are not trusted. A subset of frontbench (major) MPs from the Conservatives, Labour, and Liberal Democrats were selected as representatives of their party. These 26 MPs (across all three parties) were considered to have a greater influence on their colleagues, and had many more quotes in news media than other MPs, which aided other analysis. This simplified measure of influence tests the feasibility of the overall analytics concept. Further research work would be required to determine more complex trust relationships.

Skills data were not available directly for this study. Instead, discussion topics were used as a suitable proxy measure. A more direct measure of skill should be determined in further development of the model. Discussion topics were automatically identified from the quote data using the Empath tool. The three topics that correlated most highly with the behavioural outcome were used in the model: "breaking", "farming" and "violence".

Based on the above, the number of "observed" input variables for each MP was 16, comprising: attitude (1), emotion (6), subjective norm (1), personality (5), and skills (3). The leader and peer influence variables are binary switches that multiply the attitude and subjective norm variables. The observed output variable is the number of times $n_b$ an MP votes for Brexit out of $n_v$ possible votes.

## 6.3. Expert-specified network structure

Figure 12 shows an expert-specified structure of the BN model with plate notation. The rectangles represent repeated variables, with the number of repetitions shown above the rectangle. These are:

- $n_i = 13$. The number of intention or motivation (M) variables. These are: happiness, sadness, anger, fear, disgust, surprise, personal pro-Brexit attitude, party pro-Brexit attitude (subjective norm), openness, conscientiousness, extraversion, agreeableness, and neuroticism.
- $n_p = 26$. The number of peers in the set of opportunity (O) variables. Each peer variable represents the trust in one of the major MPs. The $+1$ term in the diagram relates to trust in the leader of the party.

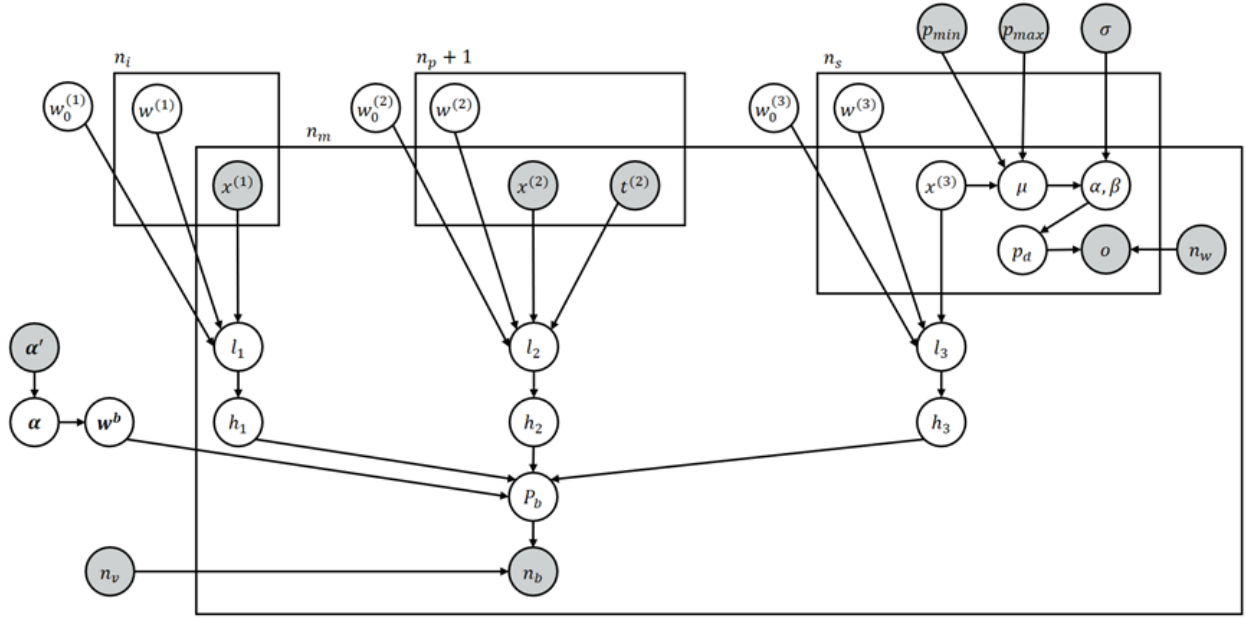

*Figure 12: Bayesian network model with plate notation for repeated variables. Shaded circles indicate observed variables.*

- $n_s = 3$. The number of skills or capability (C) variables. These relate to the "breaking", "farming", and "violence" discussion topics which are assumed to have corresponding skills associated with them.
- $n_m = 610$. The number of all MPs in the dataset.

The $w^{(\cdot)}$ variables are weights and biases for the input factors. The $x^{(\cdot)}$ variables are observed (motivation and opportunity) or unobserved (capability) features for each person. The additional variables in the top right, including the number of words spoken by each individual $n_w$, form the skills/discussion part of the model. The $\boldsymbol{\alpha}$ and $\boldsymbol{w}^b$ variables represent uncertain prior information about the relative weight of the three C/O/M branches on voting behaviour. These variables are based on $\boldsymbol{\alpha}'$, which specifies the proportions 0.787, 0.039, and 0.012, for motivation, opportunity, and capability, respectively, based on Danish voting data (Hansen *et al.*, 2008). The $l_i$ and $h_i$ variables are internal deterministic variables representing the logit and hidden probability of the three C/O/M branches affecting voting behaviour. $P_b$ is the probability of voting for Brexit, $n_v$ is the number of votes about Brexit, and $n_b$ is the number of votes for Brexit. The model structure, as with all BNs, is a directed acyclic graph (DAG). The mathematical relationship between the variables is defined in State-Davey *et al.* (2020).

The BN was built using the Python interface to the Stan tool[8]. Stan is a probabilistic programming language for statistical inference (Carpenter *et al.*, 2017). It uses a Markov Chain Monte Carlo (MCMC) method (Chib *et al.*, 1995) to estimate the uncertainty in model parameters based on measured data by generating samples from the posterior parameter distribution. The specific method used is a variation on Hamiltonian Monte Carlo (Hoffman *et al.*, 2014). Within the network, the learnt parameters of the expert-specified BN indicated that, in contrast to the prior information from Hansen *et al.* (2008), where motivation was the highest-weight hidden factor, opportunity was more important. This demonstrates the ability of BNs to supersede prior information when sufficient data are available. The overall performance of the network is discussed in relation to others in section 6.6.

**6.4. Data-driven network structures**

Behavioural theories from psychology provide direction for the data collection process and construction of data science models (Luo *et al.*, 2019). They also support interpretation of the data analysis. However, an important aspect of using data to support understanding of behaviour is that it can reveal relationships between psychosocial factors that have not been specified from behavioural theory, providing a data-driven and evidence-based approach. Learning the structure of the model directly from the data instead of specifying it from theory can enhance understanding of how factors influence one another. In this paper we explore the potential of three types of structure learning algorithms to learn the BN structure directly from training data. These are heuristic score-based, constraint-based, and hybrid types of method.

---
[8] https://mc-stan.org

Learning BNs when the network structure is not known in advance is a more complex learning problem than just learning the parameters. The number of possible structures increases super-exponentially as the number of variables (nodes) grows, and only a small fraction of the space of possible structures can be examined in a reasonable time-frame. However, heuristic search algorithms (also known as scored-based methods) or constraint-based methods can be used to focus the search on the most promising areas of the search space. Hybrid techniques combine both of these approaches to utilise the benefits of both.

We tested three algorithms from the bnlearn library[9] of the R programming language: hill climbing (HC), Peter and Clark (PC), and max-min hill-climbing (MMHC). The score-based HC algorithm starts with a "network" of unconnected nodes and, at each step in the search, adds, deletes, or reverses the direction of a directed edge connecting nodes (Scutari *et al.*, 2019). Changes are accepted in a greedy manner based on the Bayesian information criterion (BIC) score at each iteration. PC is a constraint-based algorithm. Constraint-based algorithms learn conditional independence relationships (constraints) in the training dataset (Spirtes *et al.*, 1991). We used the stable version of PC (Colombo *et al.*, 2014) and the asymptotic distribution of the t-test for hypothesis testing. MMHC is a hybrid algorithm that uses the max-min parents-and-children algorithm to learn the undirected (skeleton) structure and the HC algorithm to find the optimal directions of relationships based on the skeleton (Tsamardinos *et al.*, 2006). In the following paragraph, we describe some general qualities of the structures learnt from these algorithms. Performance comparisons are given in section 6.6.

The HC algorithm created a very dense DAG with 156 directed edges. Although the learnt model has a high likelihood, from a practical point of view, the structure is not very informative in adding to the theory as it is too dense to interpret easily and does not provide a clear understanding of latent factor structure, due to the high level of inter-relationship between factors. The PC algorithm produces a much less dense network than HC, with 40 edges. However, the structure is not a DAG as there are 35 directed and 5 undirected edges. Undirected edges occur when there is a tie in the score for the two possible directions. The presence of undirected edges prevented the library from learning BN parameters and it was therefore not possible to assess performance on the downstream task of vote prediction. The MMHC algorithm also produced a network with 40 edges with many similarities to the PC network. However all edges were directed, resulting in a valid DAG. Out of the three purely data-derived networks produced, MMHC was considered to be the most interpretable.

Network structures can be built based on a combination of both theory and data. The following section describes how this can be done by borrowing ideas from alternatives to BNs.

**6.5. Combined theory-data network structure**

BNs represent state of the art in modelling complex uncertain situations where limited data is available. Competing alternatives to BNs include multiple regression, factor analysis, and structural equation modelling. Multiple regression often refers to multiple linear regression but nonlinear versions also exist. A disadvantage of all regression techniques is that they do not include latent variables, which may be a better model of true underlying behaviour.

Factor analysis characterises the relationship between observed and unobserved variables (latent or hidden factors), which enables reduction of model dimensionality. Unlike PCA, factor analysis explicitly models the presence of error. Exploratory factor analysis (EFA) is a data-driven approach to factorizing the data. This uses data rather than theory to identify potential relationships between observed and latent variables.

One disadvantage of factor analysis is that internal structure between hidden factors is not included and may be an important element of a model. Structural equation modelling (SEM) is an extension of factor analysis that includes a structural model of relationships between latent variables. SEM is very similar to BN modelling as both techniques are structural and can characterise nonlinear directed relationships that are either pre-defined or estimated. SEM consists of a measurement model and a structural model. The measurement model characterises the relationships between latent and observed variables. The structural model characterises potential causal relationships between independent (exogenous) and dependent (endogenous) variables. In this paper we use a learnt SEM to define the structure of a BN.

The first part of building an SEM is to identify the measurement model using EFA. Several techniques exist to determine the number of factors. Here, we used the simple Kaiser criterion, which selects the number of factors that have an eigenvalue greater than unity, computed from the training data correlation matrix. This is based on the logic that only factors that explain at least the same amount of variance as a single variable are worth keeping. More advanced techniques are discussed in Preacher *et al*. (2003).

---

[9] https://cran.r-project.org/package=bnlearn

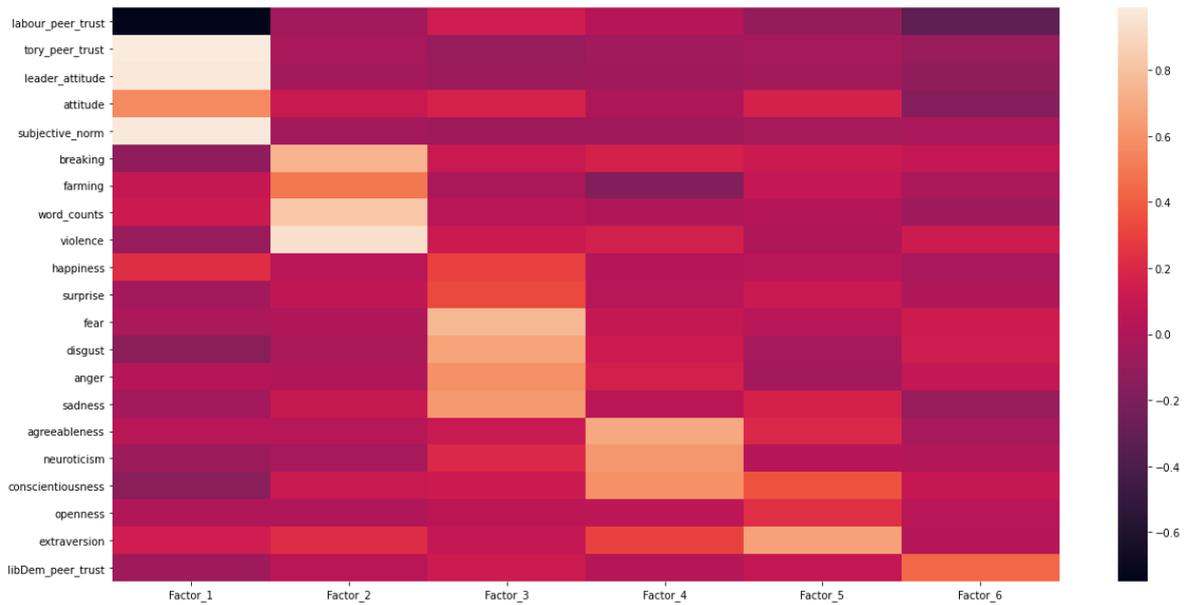

*Figure 13: Exploratory factor analysis: loadings of each measured variable against the six discovered latent factors.*

To identify the measurement model, each observed variable was assigned to the latent factor for which it had the highest absolute magnitude in the loading matrix, as shown in Figure 13. The EFA identified six latent factors relating to the psychosocial factors in the conceptual model:

- Factor 1 ("Attitude"): attitude, subjective norm, leader attitude, conservative peer trust, labour peer trust;
- Factor 2 ("Skills"): word counts, farming, violence, breaking;
- Factor 3 ("Emotion"): sadness, anger, disgust, fear, happiness, surprise;
- Factor 4 ("Personality CAN"): conscientiousness, agreeableness, neuroticism;
- Factor 5 ("Personality OE"): openness, extraversion; and
- Factor 6 ("Peer trust"): liberal democrat peer trust.

While the latent factors were obtained automatically by the algorithm, their names above have been assigned by the authors based on qualitative analysis of the observed variables assigned to them. This helps understand each factor.

In contrast to the expert-specified BN, subjective norm and leader attitude were found to load onto an "attitude" factor, along with Conservative and Labour peer trust, instead of being separate hidden factors. Only Liberal Democrat peer trust was found to load onto a "peer trust" factor. The identified "skills" and "emotion" factors match pre-specified groupings in the input variables. The personality grouping of observed variables is split into two separate factors in the SEM.

To implement the SEM the python library semopy was used (Igolkina *et al.*, 2020). Semopy takes a measurement model description, from the above EFA, and uses a fit from the training data to produce the structural model. The semopy model places links between all relevant types of variables, i.e. hidden-to-hidden, and hidden-to-output. Each link has a weight and a p-value, with the p-values indicating the statistical significance of the link. Some links are bi-directional and others are unidirectional. To convert the structure into a BN, the variables nodes must form a DAG. A number of steps were applied to the semopy structure to result in a valid BN. First, statistically insignificant links with p-values above 5% were removed. Secondly, where links are bidirectional, a single direction was selected using expert judgement. Finally, the directions of some links were reversed so that the causal relationship between variables made sense from a theoretical point of view. Since the skills factor had no statistically significant links either to other factors or the behavioural outcome node of the SEM, it was excluded from the final model.

Figure 14 shows a simplified view of the semopy structure reformatted as a BN. The input variables on the top row are the same as for the expert-specified BN (Figure 3), with the exception that trust variables are grouped by party. The middle row of unobserved variables are internal nodes. This is where the structure differs most from the expert-specified BN model, since the semopy BN model has additional hidden nodes and there is internal linkage between them. The bottom two rows of the semopy BN model (containing probability and number-of-votes nodes) are identical to the expert-specified BN model. The full DAG of this network has 26 edges, which is somewhat less than the purely data driven networks from the previous section. This network was implemented in Stan to produce the results in the following section.

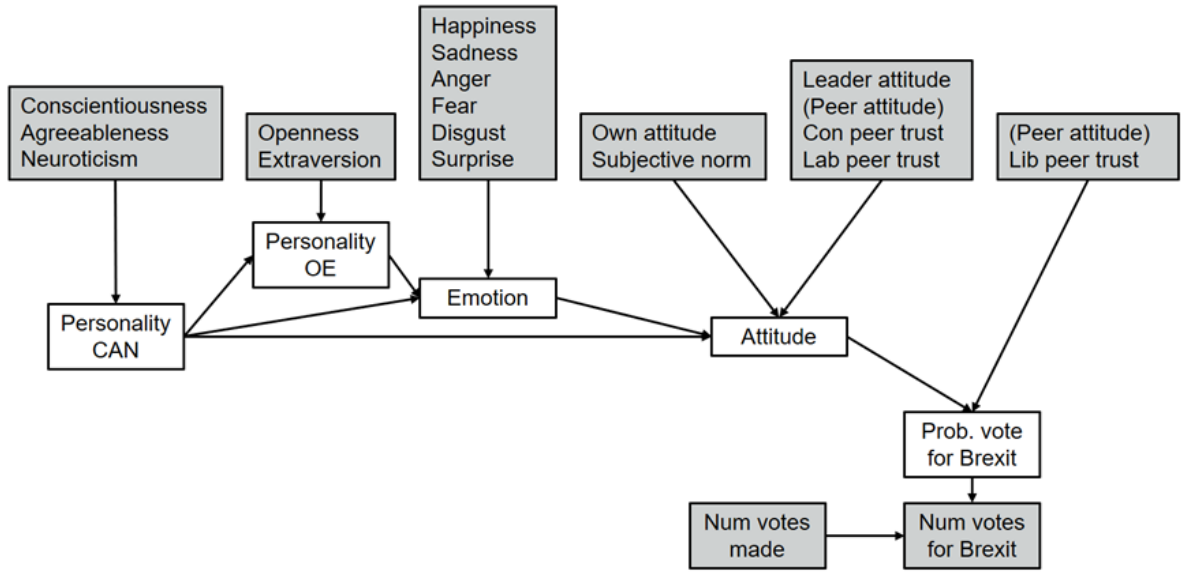

*Figure 14: Simplified view of the Bayesian network structure based on structural equation modelling.*

**6.6. Bayesian network results**

Each of the BNs described in sections 6.3 to 6.5 above were used to predict the proportion of votes for Brexit for the MPs in the Hansard test set. The main performance metric used to assess algorithms was the root mean square error (RMSE). Table 2 provides a performance comparison. Both the HC and MMHC data-driven BNs were able to outperform, by a small margin, predictions of the expert-specified BN. This indicates that, although not conclusive, the use of data-driven networks structures is a promising avenue of research. Further algorithms, performance metrics, and analyses are provided in State-Davey *et al.* (2021). A single algorithm with its default setting was chosen to represent each of the three structure learning approaches. The R bnlearn library provides other parameter settings and algorithms. These should be investigated to understand better how well a purely data driven BN model can be learnt. The hybrid approach is particularly interesting as it balances interpretability of the DAG with performance. The R bnlearn library provides the facility of allow- and deny-lists to specify edges that must or must not be present in the learnt DAG. This could be a useful option when there is partial knowledge of the domain.

| *Structure Learning Approach* | *RMSE* |
|---|---|
| *Expert-Specified* | *0.2031* |
| *Hill Climbing (HC)* | *0.1782* |
| *Max-Min HC (MMHC)* | *0.1757* |
| *SEM-derived* | *0.2241* |

*Table 2: Structure learning performance comparison for prediction of the proportion of votes in the Hansard test set.*

**6.7. Bayesian network summary**

A crucial aspect of BNs for behavioural analytics is the ability to model the relationship between the psychological factors in the model. Understanding patterns of relationship between factors and their structure is critical for intervention design and behavioural change. The expert-specified model structure was defined based on the COM-B model. However, many other behavioural change theories have been developed, such as the theory of planned behaviour (Ajzen, 1985) and social cognitive theory (Bandura, 1986), which emphasise different psychological factors and relationships between the factors. The ability to learn structure from data is therefore useful to improve the accuracies of predictions and to validate theories.

# 7. Conclusions

Behavioural analytics is the combination of psychological behavioural modelling and statistical data analysis. The key to a cutting-edge behavioural analytics capability is the development of appropriate mathematical models within a wider data processing framework, combined with behavioural theory. This enables the possibility of characterising the mathematics of the mind.

This paper has considered two strands of behavioural analytics research. The first part developed a system for tracking and predicting the descent into extremism and terrorism. Natural language processing converts arbitrary-length statements to fixed-length numerical embeddings that can be used for state-of-mind estimation and classification. The concept was verified through analysis of labelled quotes in English, French, and Arabic. The classifier detects extremism correctly 81% of the time and terrorism 97% of the time in English and 95% and 93% of the time in Arabic, respectively. Quotes in French were classified correctly as to whether or not they relate to Brexit 85% of the time. The tracking algorithms can detect trends over time and sharp changes attributed to major events. Analysis has been applied at an individual, group, and population level. The relationship between intentions, as expressed in words, and actual behavioural action has been analysed in a surrogate dataset by comparing MP voting behaviour with the statements they made on the topic of Brexit. A strong correlation was measured but, interestingly, behaviour was found to be influenced more by party membership than by individual attitude. It is anticipated that group membership in non-political contexts also influences behaviour but it has not yet been tested the extent to which this is the case.

A second aspect of this work has been to explore the development of Bayesian networks for behavioural analysis. The solutions are based on behavioural theory from the psychology literature and data-driven methods from data science, which collectively support numerical analysis of the psychological and social drivers of behaviour. The algorithms have been assessed using the MP quote and vote data, where it was found that data-driven methods can produce similar models to the expert-specified ones. The BN approach represents the state-of-the-art in modelling complex uncertain situations where limited data is available, allowing data to be combined with domain knowledge, and supporting good prediction accuracy, even with small sample sizes. These are fundamental capabilities for helping defence and security overcome some of the challenges currently faced in target audience analysis activities.

# 8. Acknowledgements

The writing of this paper was funded by the QinetiQ Fellow scheme. The original work was funded primarily through the Defence and Security Accelerator organisation. Iraq Parliament data gathering and cleaning was done by Jaber Syed. Arabic language analysis was performed by Andrew Snell. Translation and French name normalisation was done by Chris Hutber. Bayesian network and analysis processing pipelines were implemented by Stuart Bertram. Internet searches and data collation were carried out using a bespoke semi-automated tool developed by Thomas Lane Cassell. This paper is copyright © QinetiQ 2023.